\documentclass[aps,onecolumn,superscriptaddress,showkeys,showpacs,longbibliography,nofootinbib]{revtex4-2}
\usepackage{graphicx}
\usepackage{bm}
\usepackage{graphicx}
\usepackage{amsmath}
\usepackage{mathrsfs}
\usepackage{ulem}
\usepackage{color}
\usepackage[colorlinks, linkcolor=red,anchorcolor=green,citecolor=blue]{hyperref}

\graphicspath{{./figs/}}

\newcommand{\nn}{\nonumber}
\newcommand\diag{\operatorname{diag}}

\begin{document}  
	
\title{Propagation of spin channel waves }

\author{Jin Hu}
\email{hu-j17@mails.tsinghua.edu.cn}
\author{Zhe Xu}
\affiliation{Department of Physics, Tsinghua University, Beijing 100084, China}

\begin{abstract}
 A mutilated model  is constructed  to approximate the collision term of  spin Boltzmann equation that  incorporates newly appearing collisional invariants i.e, the total angular momentum.  With recourse to degenerate perturbation theory, the dispersion relations of hydrodynamic modes are formulated, among which spin modes are responsible for spin equilibration. We find that the non-locality does not change the sound speed but slows down the propagation of spin channel waves. The damping rates of spin modes are close to those of spinless modes over a reasonable  parameter value range. The results reveal that both spin and momentum should be treated simultaneously in a unified transport framework.    In the nonrelativistic limit,  the short-wavelength behavior for normal modes is also explored  and  there exists a critical point for every distinct discrete  mode over which only quasiparticle modes contribute.

\end{abstract}

%\preprint{USTC-ICTS/PCFT-20-13}

%\pacs{25.75.-q, 25.75.Cj, 25.75.Ld, 47.75+f}

% 25.75.Ld Collective flow
% 24.10.Nz Hydrodynamic models
% 47.75.+f Relativistic fluid dynamics
% 25.75.-q Relativistic heavy-ion collisions
% 25.75.Cj Leptons production in relativistic heavy-ion collisions

\maketitle	

\section{Introduction}
The  experimental developments in measuring the spin-related observables of $ \Lambda $ hyperons \cite{STAR:2017ckg,Alpatov:2020iev,Adam:2019srw,Adam:2018ivw} have raised extensive interests in global polarization
\cite{Liang:2004ph,Wei:2018zfb,Karpenko:2016jyx,Csernai:2018yok,Bzdak:2017shg,Shi:2017wpk,Sun:2017xhx,Ivanov:2019wzg,Xie:2017upb} and  local polarization \cite{Becattini:2017gcx,Xia:2018tes}. Theoretical calculations are not consistent with experiment results in local polarization, which is referred to as ``spin sign problem'' and originates from the fact that spin does not reach equilibrium as expected, see \cite{Becattini:2022zvf} for a recent review.  In the past few years, different efforts are made  to get insight into the spin puzzle along the line of  spin hydrodynamics \cite{Florkowski:2017ruc,Peng:2021ago,Becattini:2009wh, Florkowski:2018fap,Hattori:2019lfp,Fukushima:2020ucl,Hu:2021lnx,Bhadury:2020cop,Hu:2021pwh,Hu:2022lpi,Hu:2022xjn,Hu:2022azy,Weickgenannt:2022zxs}, and quantum kinetic theory \cite{Weickgenannt:2021cuo,Weickgenannt:2020aaf,Yang:2020hri,Sheng:2021kfc,Chen:2021azy,Wang:2021qnt,Yang:2021fea}, among which spin relaxation becomes essential because the relaxation rate of the spin density toward its equilibrium value is crucial in determining how the spin polarization evolves in time in theoretical simulations of QCD plasma. 
%However, the resulting equations are usually too involved to be solved at acceptable costs, which calls for reasonable approximation to obtain the results closely related to experiments. Therefore recently,  spin relaxation time has triggered lots of attention, which could provide phenomenological constraints in the construction of  spin kinetic theories. 

When we talk about spin equilibration  of a spinful system, there is a increase of degrees of freedom compared to a spinless system. Considering a collection of quasi-particles with nonzero spins, the dynamic variables are necessarily enlarged to include another particle property spin, i.e, the distribution function for describing the system made of quasi-particles has dependence on spin. From a more general view irrespective of quasi-particle picture, the evolution functions turn out to be conservation laws
$\partial_\mu T^{\mu\nu} =0, \,\partial_\lambda \Sigma^{\lambda\mu\nu} =0$ where $T$ and $\Sigma$ denote the energy momentum tensor and total angular momentum tensor. Canonically,  we allow a division $ \Sigma^{\mu\alpha\beta}
\equiv (x^\alpha T^{\mu\beta}  - x^\beta T^{\mu\alpha} ) +  S^{\mu\alpha\beta} $, where $x$ is the space-time coordinate and  $ S^{\mu\alpha\beta} = -S^{\mu\beta\alpha} $ is newly defined spin tensor.
In this context, the local equilibrium state is defined as the state with maximum local entropy or vanishing divergence of local entropy four-current. For a spinful system, the entropy current receives the contribution from spin tensor, which is exactly reflected in the form of statistical operator \cite{Becattini:2018duy,Hu:2021lnx,Hu:2022azy}.

Similar to the still unsettled question of how thermal equilibrium in relativistic heavy-ion collisions is reached,  the questions of how the spin of quarks relaxes to equilibrium and whether it equilibrates faster than momentum or not remain under debates. Note that there have been many efforts exploring the spin relaxation rate   including but not limited to  perturbative QCD techniques \cite{Li:2019qkf,Kapusta:2019sad,Kapusta:2020npk,Hongo:2022izs}, the Nambu–Jona- Lasinio model \cite{Kapusta:2019ktm}, the AdS/CFT correspondence \cite{Li:2018srq}, an effective vertex for the interaction with the thermal vorticity \cite{Ayala:2019iin,Ayala:2020ndx} and a second-order spin hydrodynamic calculation \cite{Weickgenannt:2022zxs}.
%In our constructed framework, one can carefully adjust parameters to model various physical scenarios with distinct interactions,  
When spin is considered, one must take into account the conservation of total angular momentum in hydrodynamic description especially for a system with considerable spin-orbit conversion due to frequent interactions. Accompanied by newly introduced degrees of freedom, there arises new physical phenomenon of the propagation of spin channel waves. Analogous to sound propagation in spinless (spin-averaged) fluids \cite{DeGroot:1980dk}, the propagation of spin channel waves should be also fundamental in spin hydrodynamic theory and deserve a comprehensive exploration, which we expect to provide in this work.

As with the recent studies \cite{Hongo:2021ona,Hongo:2022izs,Hu:2022lpi},   we  focus on a linear  mode analysis in the current work and find that spin relaxation couples to  the attenuation of spin modes. Therefore,  spin relaxation time is identified  as the lifetime of spin modes, based on which one can make a direct comparison of two typical time scales in relation with spin and momentum relaxation. On the other hand, the researches on spin polarization directly promotes the developments of spin kinetic theory. Once spin is considered,  there also arises new physical phenomenon of the propagation of spin channel waves. Analogous to sound propagation in ordinary fluids, the propagation of spin channel waves should be also fundamental in spin hydrodynamic theory. 
%Therefore, we provide  a comprehensive exploration about it.

This paper is organized as follows. In Sec.~\ref{lated}
we show how to construct  the  mutilated collision term incorporating six new collision invariants introduced by spin Boltzmann equation \cite{Weickgenannt:2021cuo}. In Sec.~\ref{expansion} we adopt the  method of degenerate perturbation theory widely used in quantum mechanics to derive  the dispersion relations of discrete normal modes  up to second order in wavenumbers where the discussion of spin relaxation is also presented. After that, the comparison with other associated research works  is given in Sec.~\ref{cop}. In Sec.~\ref{non}, we talk about the short wavelength behavior of those normal modes. Summary and outlook are given in Sec.~\ref{su}.
Natural units $k_B=c=\hbar=1$ are used. The metric tensor is given by $g^{\mu\nu}=\diag(1,-1,-1,-1)$ , while $\Delta^{\mu\nu} \equiv g^{\mu\nu}-u^\mu u^\nu$ is the projection tensor orthogonal to the four-vector fluid velocity $u^\mu$ and  $\epsilon^{\mu\nu\alpha\beta}$ is Levicivita tensor.
In addition, we employ the antisymmetric shorthand,
$X^{[ \mu\nu ] } \equiv (X^{ \mu\nu } - X^{ \nu \mu})/2.$ 
%For  more details, e.g. various matrix elements $V^{(\lambda)}_{i,j},H^{(\lambda)}_{i,j}$ and $Q^{(\lambda)}_{i,j}$ appearing in the main text, we refer to \cite{Hu:2022lpi}.

 \section{Mutilated collision operator}
 \label{lated}
When it comes to spin-induced phenomena, the framework of transport theory must be extended to incorporate the information of spin evolution.  
%As a long-wavelength effective theory, spin hydrodynamic theory itself can't provide the details of dynamics, for instance, the specific value of all transport coefficients, which should be derived from microscopic transport theory. 
To that end, the spin Boltzmann equation for massive spin-$1/2$ fermions with non-local collision effects was proposed, where the consistent interpretation for equilibrium state and collision invariants is elaborated \cite{Weickgenannt:2021cuo}.

 Here we choose to work out the propagation of hydrodynamic modes by constructing a mutilated collision operator based on the collision invariants appearing in \cite{Weickgenannt:2021cuo} instead of directly solving the complicated linearized integral equation derived in \cite{Hu:2022lpi}.
Note that the collision invariants are exactly  eigenfunctions of linearized collision operator with zero as  eigenvalues. If insisting on semipositive definiteness and self-adjointness of a linearized collision operator, we are led to treat it as an evolution Hamiltonian operator \cite{Hu:2022lpi}
\begin{align}
\label{boltz}
p\cdot \partial \chi(x,p,\bm{s})=-L\chi(x,p,\bm{s}),
\end{align}	 
where we assume that there is no external field in the above linearized transport equation.
 In addition, $\chi(x,p,\bm{s})$ represents the  deviation function from equilibrium distribution dependent on space-time coordination $x$, particle momentum $p$ and spin $\bm{s}$, and $L$ denotes the linearized collision operator with its explicit form temporarily uncovered.
 
 Under homogeneous circumstance, Eq.(\ref{boltz}) can be formally solved and the resulting solution is
  \begin{align}
  \label{evo}
  \chi(t,p,\bm{s})=\exp(-\frac{L}{u\cdot p}t)\chi(p,\bm{s}),
  \end{align}
  with the initial condition $\chi(t=0,p,\bm{s})=\chi(p,\bm{s})$.
  
  According to Hamiltonian formalism, we can always expand the deviation function as a sum of linear superposition of the eigen states of linearized collision operator $L$. As time goes by, zero modes can survive long time while positive ones become damped exponentially and less dominant. Hereafter, we concentrate on the  zero modes and see how they respond to the perturbation of non-uniformity, namely, the space coordination dependence of  deviation function $\chi(x,p,\bm{s})$  is recovered.

%solve the transport problem in the manner of quantum mechanics,  while the non-negativity is rather necessary for linear analysis because positive eigen spectrum leads to instability.   

%Additionally, a quiescent background fluid i.e., $u^{\mu}=(1,0,0,0)$ is chosen in the remainder of this paper.
  The linearized collision operator $L$ is now approximated by a mutilated operator
\begin{align}
\label{L}
-L\sim \big(-\gamma+\gamma\sum_{n=1}^{11}|\lambda_n\rangle\langle \lambda_n|\big)
\end{align}
 with $|\lambda_n\rangle$ being orthonormal eigenfunctions of $L$  and $\gamma$ being a representative positive eigenvalue. One can easily  verify that $L$ inherit basic properties of what we require for a linearized collision operator such as semipositive definiteness, self-adjointness and $L|\lambda_n\rangle=0, n=1,\cdots 11$ and $L|\lambda_n\rangle=\gamma |\lambda_n\rangle, n>11$. Here the ``mutilated" means all  positive eigenvalues collapse into one chosen positive eigenvalue (it is suggestive to take the smallest one). The zero modes with eleven-fold degeneracy exactly correspond to collision invariants $1,p^\mu$ and  $J^{\mu\nu}$, where  the identification of $J^{\mu\nu}\equiv2\Delta^{[\mu}p^{\nu]}+\frac{1}{2}\Sigma^{\mu\nu}_{\bm{s}}$ as total angular momentum is seen in \cite{Weickgenannt:2021cuo,Weickgenannt:2020aaf} with
 \begin{align}
 \Delta^{\mu}\equiv-\frac{1}{2m(p\cdot\hat{t}+m)}\, 
 \epsilon^{\mu\nu\alpha\beta}p_\nu \hat{t}_\alpha \bm{s}_{\beta}
 \end{align}
 characterizing the non-locality in a collision, where $\hat{t}^\mu$ is the time-like unit vector which is $(1,\boldsymbol{0})$ 
 in the frame where $p^\mu$ is measured. 
 
  It may be observed that this is exactly  a kind of relaxation time approximation (RTA) by identifying $\gamma$ with the reciprocal of relaxation time $\tau_R^{-1}$. Compared to traditional RTA, the novel RTA or mutilated operator is proved to reconcile the momentum dependence of the relaxation time with the macroscopic conservation laws. 
  When the relaxation time has no momentum dependence, one can always argue that RTA is consistent with the conservation laws by imposing matching conditions but this is not the general case. Without elaboration, we refer to a recent letter \cite{Rocha:2021zcw}. 
    From now on,  $\gamma$ is parameterized as energy dependent $\gamma\equiv\gamma_R/\tau^\lambda$ with  an energy-independent constant $\gamma_R$. 
  
  In order to seek a solution of the form, $\chi\sim\tilde{\chi}e^{-ik\cdot x}$, we substitute it into Eq.(\ref{boltz}), 
    \begin{align}
    \label{cf}
    &\tau \omega\tilde{\chi}+\hat{p}^\mu l_\mu\kappa\tilde{\chi}=-iL\tilde{\chi},
    \end{align}
    where the linearized collision operator is put less abstract with the specified form shown as
    \begin{align}
   L\tilde{\chi}\equiv\gamma\tau\big(\tilde{\chi}-\sum_{n=1}^{11}(\tilde{\psi}_n,\tau\tilde{\chi})\tilde{\psi}_n\;\big),
    \end{align} 
 with notations $\tau\equiv \frac{p\cdot u}{T},\hat{p}\equiv\frac{p}{T}$. Here we introduce dimensionless frequency and wave vector
 \begin{align}
 \label{pert}
  \omega\equiv \frac{u\cdot k}{n\sigma},\quad \kappa^\alpha\equiv\frac{\Delta^{\alpha\beta}k_\beta}{n\sigma},
 \end{align}
 and one unit vector
  \begin{align}
  \label{pert2}
 l^\alpha\equiv\frac{\kappa^\alpha}{\kappa},\quad \kappa\equiv\sqrt{-\kappa\cdot\kappa},
  \end{align}
 where  $n$ is density, $T$ is temperature and $\sigma$ is an arbitrary constant with the dimension of cross sections.
 Additionally, the inner product  is defined as
 \begin{align}
 \label{inner}
 (B,C)=\frac{1}{(2\pi)^3}\int d\Gamma \exp(-\beta\cdot p)B(p,\bm{s}) C(p,\bm{s}),
 \end{align}
with the  measure  defined as $d\Gamma \equiv d^4p\, \delta(p^2 - m^2) \sqrt{\frac{p^2}{3 \pi^2}} d^4\bm{s}\, \delta(\bm{s}\cdot\bm{s}+3)
\delta(p\cdot \bm{s})$ \cite{Florkowski:2018fap},
 and the eigenfunctions $|\lambda_n\rangle$ are also replaced by less-abstract functions $\tilde{\psi}_n$ given by
  \begin{align}
  \label{set}
  &\tilde{\psi}_{1}=\frac{1}{\sqrt{V^{(0)}_{1,1}}}, \quad \tilde{\psi}_{2}=\beta\frac{u\cdot p-\frac{I_{20}}{I_{10}}}{\sqrt{V^{(0)}_{2,2}}},\quad \tilde{\psi}_{3}=\frac{\beta l\cdot p}{\sqrt{V^{(0)}_{3,3}}},\nn\\
  &\tilde{\psi}_{4}=\frac{\beta j\cdot p}{\sqrt{V^{(0)}_{3,3}}}, \quad \tilde{\psi}_{5}=\frac{\beta v\cdot p}{\sqrt{V^{(0)}_{3,3}}},\quad
  \tilde{\psi}_{6}=\frac{u_\mu J^{\mu\nu}l_\nu}{\sqrt{V^{(0)}_{6,6}}},\nn\\ &\tilde{\psi}_{7}=\frac{u_\mu J^{\mu\nu}j_\nu}{\sqrt{V^{(0)}_{6,6}}},\quad \tilde{\psi}_{8}=\frac{u_\mu J^{\mu\nu}h_\nu}{\sqrt{V^{(0)}_{6,6}}},\quad \tilde{\psi}_{9}=\frac{l_\mu J^{\mu\nu}j_\nu}{\sqrt{V^{(0)}_{9,9}}},\nn\\
  &\tilde{\psi}_{10}=\frac{l_\mu J^{\mu\nu}h_\nu}{\sqrt{V^{(0)}_{9,9}}}, \quad \tilde{\psi}_{11}=\frac{j_\mu J^{\mu\nu}h_\nu}{\sqrt{V^{(0)}_{9,9}}},
  \end{align}
  where
  \begin{align}
  I_{nq} &\equiv \frac{2}{(2q+1)!!} \int \frac{\rm d^4p}{(2\pi)^3}\,\delta(p^2 - m^2)(u\cdot p)^{n-2q},\\
  V^{(\lambda)}_{i,j}&\equiv\beta\int \frac{\rm d\Gamma}{(2\pi)^3}\tau^\lambda\tilde{\psi}_i\kappa\cdot p\,\tilde{\psi}_j\exp(-\beta \cdot p).
  \end{align}
 Here $\beta^\mu\equiv\beta u^\mu=\frac{u^\mu}{T}$ and we  introduced two auxiliary unit vectors $j, h$ to form an orthonormal triad with $u$ and $l$. In the remainder of this work, a quiescent background fluid i.e., $u^{\mu}=(1,0,0,0)$ is chosen, then the triad $(u,j,h,l)$ stands for the projection to the directions of $(t,x,y,z)$.  Last but not the least, note the eigenfunctions Eq.(\ref{set}) has been defined to fulfill the orthonormal condition
 \begin{align}
 \label{one}
 (\tilde{\psi}_\alpha,\tau \tilde{\psi}_\beta)=\delta_{\alpha\beta}.
 \end{align}

 \section{Degenerate perturbation theory}
 \label{expansion}
 As a familiar problem in the perturbation theory,  the solutions to Eq.(\ref{cf}) can be sought in the fashion as used in quantum mechanics by treating the spatial term $\hat{p}^\mu \kappa_\mu \tilde{\chi}$ as a perturbation with respect to  $-iL\tilde{\chi}$, then the eigenfunctions and eigenvalues can be routinely  expanded into
  \begin{align}
  &\tilde{\chi}=\tilde{\chi}^{(0)}+\tilde{\chi}^{(1)}+\cdots,\nn\\
  &\omega=\omega^{(0)}+\omega^{(1)}+\omega^{(2)}+\cdots.
  \end{align}
  
The dispersion relations, which are obtained from the secular equation for Eq.(\ref{cf}), are formulated up to first order in $\kappa$ \cite{Hu:2022lpi}
 \begin{align}
 &\omega^{(1)}_1=-\omega^{(1)}_2=\sqrt{H^{(0)2}_{2,3}+H^{(0)2}_{1,3}}\nn\\ &\omega^{(1)}_{3}=\omega^{(1)}_{4}=\omega^{(1)}_{5}=\omega^{(1)}_6=\omega^{(1)}_{11}=0,\nn\\
 &\omega^{(1)}_7=\omega^{(1)}_8=H^{(0)}_{7,9},\quad \omega^{(1)}_{9}=\omega^{(1)}_{10}=-H^{(0)}_{7,9},
 \end{align}
 where
 \begin{align}
 H^{(\lambda)}_{i,j}\equiv\beta\int \frac{\rm d\Gamma}{(2\pi)^3}\tau^\lambda\tilde{\psi}_i\kappa\cdot p\,\tilde{\psi}_j\exp(-\beta \cdot p).
 \end{align}	
 
\begin{figure}[!htb]
 	\includegraphics[width=0.5\textwidth]{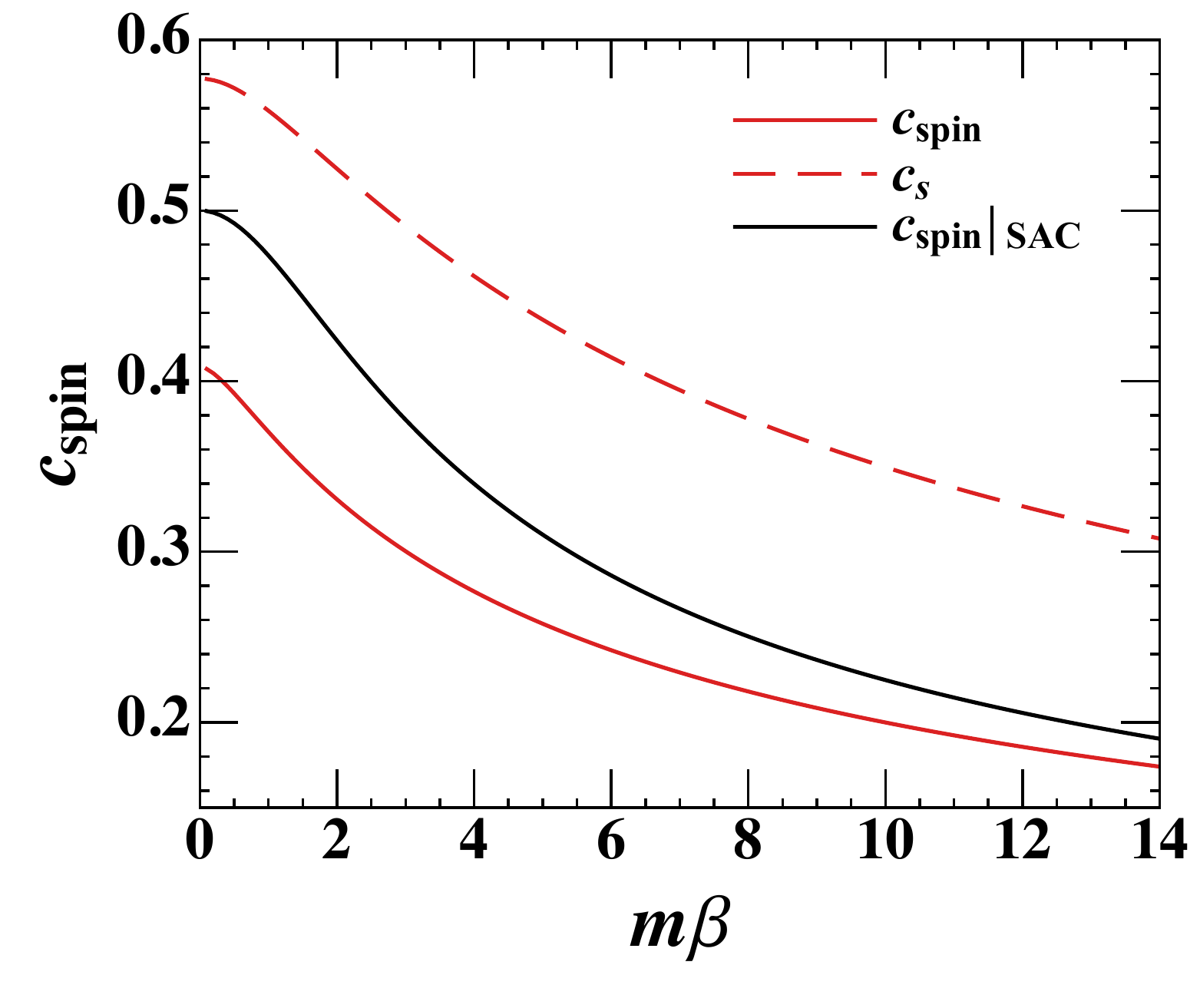} \ ~ \ ~  
 	\caption{The propagating speed of discrete normal modes as a function of $\beta m$.}
 	\label{fig1}
 \end{figure}

  One can readily verify that the results of the first five spinless modes are the same as those in \cite{DeGroot:1980dk} independent of the details of interactions involved. At the meantime, we find that only four transverse spin modes ($\omega_7,\omega_8,\omega_9,\omega_{10}$) are propagating with the same speed of propagation $c_{\text{spin}}=\omega^{(1)}_7/\kappa$ where the minus sign represents the opposite traveling direction. The dependence of propagating velocity of discrete normal modes on the reduced mass $z$ ($z\equiv\beta m$) is exhibited in Fig.\ref{fig1}. As a comparison, we also exhibit the sound speed $c_s=\omega^{(1)}_1/\kappa$,  which is larger than $c_{\text{spin}}$. In addition, $c_{\text{spin}}$  under the condition of spin angular momentum conservation (SAC) is  shown by the solid black line. When setting the non-locality $\Delta$ zero, $c_{\text{spin}}$  returns to $c_{\text{spin}}|_{\text{SAC}}$   that is also presented in \cite{Ambrus:2022yzz}. 
  
  It can be  seen clearly from Fig.\ref{fig1}  that the propagating speeds of sound modes and transverse spin modes are monotonically decreasing functions of $z$ and the inclusion of  non-locality in collisions  will slow down the propagation of spin channel waves (we call them spin channel waves in analogy with sound waves), while the propagation of sound wave is immune to that change. The non-locality does not alter the equation of state (EOS) in contrast to Enskog-type non-locality with hard-sphere exclusion potential. A quick inspection shows that $1$ and $p^\mu$ are not  collision invariants of Enskog collision term any more and the redefinition or modification to the static pressure is needed to retain the form of conservation laws \cite{Malfliet:1984ix}.
  
 Up to the second order in $\kappa$, the dispersion relations or frequencies are summarized as
 \begin{align}
 \label{frequency}
 & \omega_1=c_s\kappa-i\Gamma_1, \quad\omega_2=-c_s\kappa-i\Gamma_1,\nn\\ &\omega_3=-i\Gamma_3,\quad  \omega_4=\omega_5=-i\Gamma_4,\quad 
  \nn\\
 &\omega_6=-i\Gamma_6,  \quad\omega_7=\omega_8=c_{\text{spin}}\kappa-i\Gamma_7,\nn\\
 &\omega_{9}=\omega_{10}=-c_{\text{spin}}\kappa-i\Gamma_7,\quad\omega_{11}=-i\Gamma_{11},
 \end{align}
 with the damping coefficients defined as
 \begin{align}
 \Gamma_1&\equiv\frac{1}{\gamma_R}\Big[Q^{(\lambda)}_{1,1}+\frac{1}{2}\omega^{(1)2}_1\big(\frac{H^{(0)2}_{1,3}}{\omega^{(1)2}_1}\frac{V^{(\lambda)}_{1,1}}{V^{(0)}_{1,1}}+\frac{H^{(0)2}_{2,3}}{\omega^{(1)2}_1}\frac{V^{(\lambda)}_{2,2}}{V^{(0)}_{2,2}}\nn\\
 &+\frac{V^{(\lambda)}_{3,3}}{V^{(0)}_{3,3}}+\frac{2H^{(0)}_{1,3}H^{(0)}_{2,3}V^{(\lambda)}_{1,2}}{\omega^{(1)2}_1\sqrt{V^{(0)}_{11}V^{(0)}_{2,2}}}\big)-2\omega^{(1)}_1\big(\frac{H^{(0)}_{1,3}}{\omega^{(1)}_1}H^{(\lambda)}_{1,3}\nn\\
 &+\frac{H^{(0)}_{2,3}}{\omega^{(1)}_1}H^{(\lambda)}_{2,3}\big)\Big],\nn\\
 \Gamma_3&\equiv\frac{1}{\gamma_R}Q^{(\lambda)}_{3,3},\quad  \Gamma_4\equiv\frac{1}{\gamma_R}Q^{(\lambda)}_{4,4},\quad 
 \Gamma_6\equiv\frac{1}{\gamma_R}Q^{(\lambda)}_{6,6},\nn\\
 \Gamma_7&\equiv\frac{1}{\gamma_R}\big[Q^{(\lambda)}_{7,7}+\frac{1}{2}H^{(0)2}_{7,9}\big(\frac{V^{(\lambda)}_{6,6}}{V^{(0)}_{6,6}}+\frac{V^{(\lambda)}_{9,9}}{V^{(0)}_{9,9}}\big)-2H^{(0)}_{7,9}H^{(\lambda)}_{7,9}\,\big],\nn\\
 \Gamma_{11}&\equiv\frac{1}{\gamma_R}Q^{(\lambda)}_{11,11},
 \end{align}
 where
\begin{align}
Q^{(\lambda)}_{i,j}&=\beta\int \frac{\rm d\Gamma}{(2\pi)^3}\tilde{\chi}_i\frac{\tau^\lambda(p\cdot \kappa)^2}{p\cdot u} \tilde{\chi}_j\exp(-\beta \cdot p),
\end{align}	
and $\tilde{\chi}_i $'s are given in Eq.(52) of \cite{Hu:2022lpi}.

\begin{figure}[!htb]
  	\includegraphics[width=0.8\textwidth]{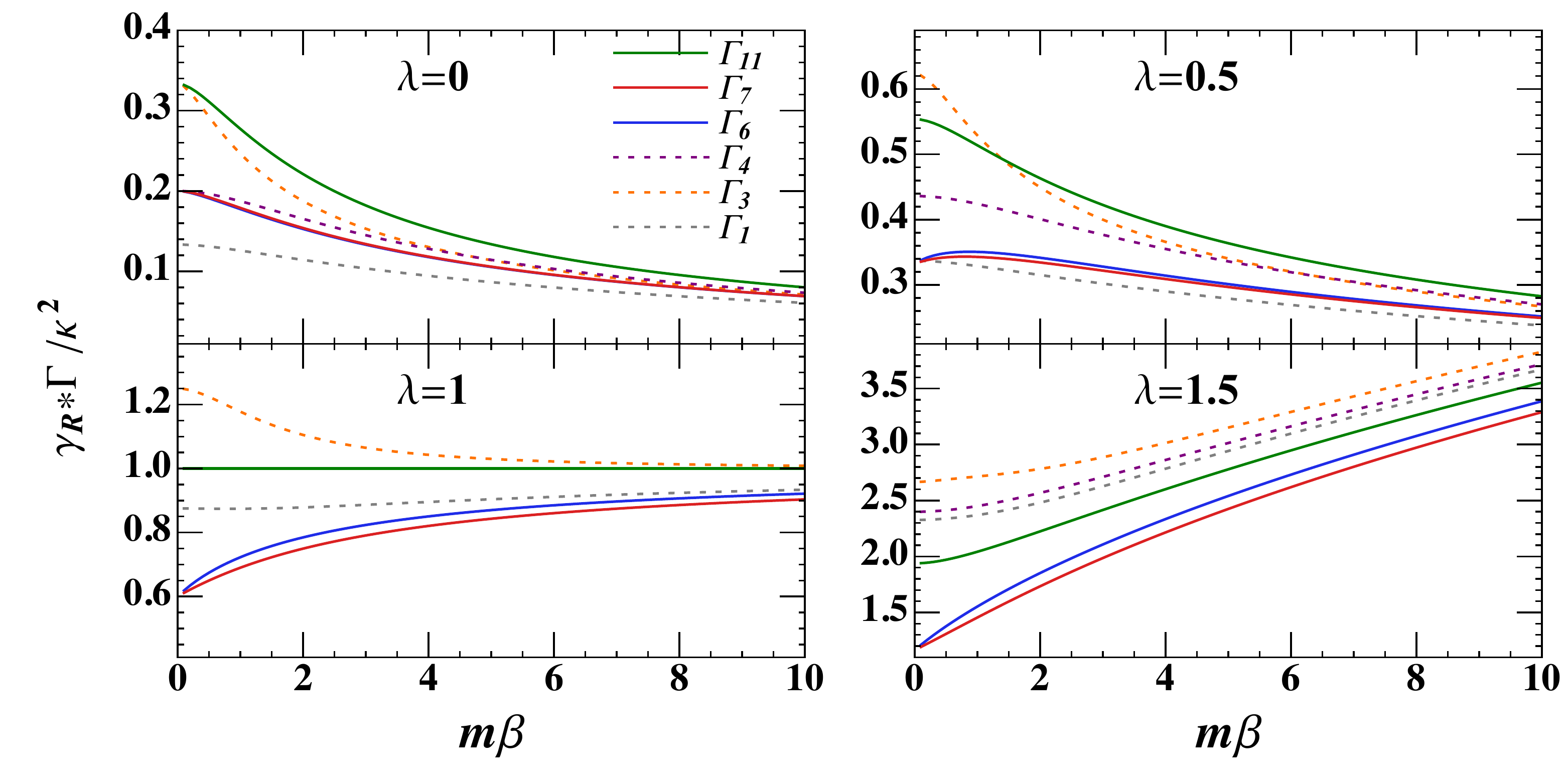} \ ~ \ ~  
  	\caption{The damping coefficients of discrete normal modes as functions of $\beta m$. The spin modes are denoted by solid lines while the spinless modes are denoted by dashed lines}
  	\label{fig2}
  \end{figure}

 Then all the damping rates are shown in Fig.\ref{fig2} as functions of reduced  mass $z$,
  which are crucial because the attenuation of spin modes couples to the dissipation of spin density \cite{Hu:2022lpi} while other spinless modes are not responsible for it.
 Among the spin modes  four propagating transverse modes are degenerate in the damping rates $\Gamma_{7}$, while the other two are non-propagating longitudinal modes which are purely decaying at their respective decaying rates $\Gamma_6$ and $\Gamma_{11}$. One can clearly see that the choice of $\lambda$ dramatically affects the attenuation of these discrete normal modes both in magnitude and in their dependence on reduced mass $z$. 
 
  The specific value of $\lambda$ relies on the dynamic details and  corresponds to various physical scenarios. For example, $\lambda=0$ corresponds to traditional RTA proposed firstly by Anderson and Witting (AW) \cite{1974Phy....74..466A}, while $\lambda=0.5$ is argued to well approximate the effective kinetic descriptions of quantum chromodynamics \cite{Rocha:2021zcw,Dusling:2009df,Dusling:2011fd,Kurkela:2017xis}. \footnote{Strictly speaking, $\lambda=0.38$ is the best fit for QCD scenario but here we take a comparatively close value $0.5$. Besides, $\lambda=0$ and $\lambda=1$ are thought to be two extreme limits between which most theories lie. There are also exceptional cases such as four fermions interaction in the electroweak sector where the energy scale is far below the  masses of gauge Bosons. In that case, $\sigma$ is shown to be proportional to energy square $E^2$ which gives an estimation of $\lambda\approx -1$ and is not within our range of consideration see \cite{Dusling:2009df,Dusling:2011fd} for more details.}
  When $\lambda$ is  big enough (see $\lambda=1.5$) and may be out of realistic range $0\lesssim \lambda\lesssim1$, the tendency even flips compared to AW case, all damping coefficients are monotonically increasing functions of $z$, and the spin modes are separated from spinless ones forming the hierarchy of $\Gamma_{\text{spin}}\lesssim\, \Gamma_{\text{nonspin}}$.   On the other hand, when $\lambda$ is comparatively small $\lambda=0$ or corresponds to relevant QCD scenario $\lambda=0.5$, the damping is almost as slow as spinless ones, namely, the dynamic evolution of spin and momentum are twisted in this scenario over a wide range of $z$. As a short summary, we find that  there is no obvious separation of two different relaxation time scales for discrete normal modes over a relevant value range of $0\lesssim \lambda\lesssim1$, which reveals the necessity of  unified transport of spin and momentum. Therefore, we should treat the evolution of both spin and momentum on the same footing.
  %In addition, there is an another minor observation that in both sectors, the hierarchy of various damping rates belonging to discrete normal modes is not influenced by the variation of $\lambda$, which is easily to be seen that the order in magnitude of solid lines or dashed lines does not vary.
  
 \section{Compare with other theoretic results}
 \label{cop}
 In this section, we provide a detailed comparison of current work with other related researches \cite{Ambrus:2022yzz,Hattori:2019lfp,Hongo:2021ona}. They have something in common and also many differences, which can be elaborated in three aspects.
 \begin{itemize}
 	\item
 	Framework. In mentioned researched works, the authors all first construct a first-order hydrodynamics based on thermodynamic second-law \cite{Hattori:2019lfp,Hongo:2021ona} and AW relaxation time approximation \cite{Ambrus:2022yzz}. After that, a linearization around the ideal fluids is made to give the analysis on the hydrodynamic modes. Wheres, one can make it directly via transport equation without recourse to hydrodynamics.
 	%thanks to the job of G.E.Uhlenbeck and C.S.Wang Chang \cite{chang}. 
 	Here we take spin Boltzmann equation as the start point and concentrate on the analysis on zero kinetic modes, which forms one-to-one correspondence to hydrodynamic modes in the limit of long wavelength \cite{Balescu}. In this sense, they are commonly called discrete normal modes. However, we have to admit that the current formalism only concerns with what is conserved, i.e, hydrodynamic modes.
 	\item
 	Degrees of freedom.	This aspect closely concerns the definition of total angular momentum tensor $J^{\lambda,\mu\nu}$. Note that almost all spin hydrodynamics on the market take the same definition of $J^{\lambda,\mu\nu}$ as what we employ in this paper. Because of the antisymmetry of last two indices, the independent degrees of freedom are six, two longitudinal ones and four transverse ones, which holds no matter what pseudo-gauge specifying the energy-momentum tensor and spin tensor is taken. However, recently in \cite{Hongo:2021ona}, the authors recommend  a new definition for $J^{\lambda,\mu\nu}$: $J^{\lambda,\mu\nu}$ is now totally antisymmetric in all indices. Other than widely used traditional definitions,  the degrees of freedom reduce to only three, which excludes three degrees of freedom related to boost symmetry. 
 	\item
 	The existence of propagating degrees of freedom. The calculations in \cite{Ambrus:2022yzz} exactly correspond to the case of $\lambda=0$ where the conservation of spin angular momentum is assumed. If turning off the nonlocality, namely, setting the nonlocal collision shift $\Delta^\mu$ zero,  our results will return to that of \cite{Ambrus:2022yzz}. Protected by the conservation laws, these two works all report that only transverse spin modes propagate similar to the behavior of electromagnetic wave but contrary to longitudinal sound modes. While in related studies \cite{Hattori:2019lfp,Hongo:2021ona}, the authors concentrate on non-conserved spin density, which is inherently relaxation-type. Therefore,  there are no propagating degrees of freedom in spin sector therein. 
 	As a supplement, we note that  if we naively construct first order theory of spin hydrodynamics via gradients expansion, spin angular momentum is  still conserved even if the nonlocality of collisions is taken into account \cite{Hu:2021pwh}, then the linear mode analysis based on hydrodynamic equations does not alter compared with  \cite{Ambrus:2022yzz}. One should expect something new emanates at the second order. 
    
 \end{itemize}

 \section{The  behavior of hydrodynamic modes in nonrelativistic limit}
 \label{non}
In this section, the existence of spin modes and their behavior outside the range of perturbation  are explored. For simplicity, we take the nonrelativistic limit and neglect the energy dependence of $\gamma$. In the nonrelativistic limit $p^\mu\rightarrow m(1,\bm{v})$ with the particle three velocity $\bm{v}$, which implies   $d\Gamma\rightarrow\frac{m^2}{2\sqrt{3}\pi}\int d^3vd^3\bm{s}\delta(\bm{s}^2-3)$,
  and
 \begin{align}
 \label{cf61}
 & \sqrt{\frac{z}{2}}(-i\omega\tilde{\chi}+i\bm{\kappa}\cdot \bm{v} \tilde{\chi})=-\gamma\big(\tilde{\chi}-\sum_{n=1}^{11}(\tilde{\psi}_n,\tilde{\chi})\tilde{\psi}_n\;\big),
 \end{align}
 where a factor $\sqrt{\frac{z}{2}}$ has been absorbed into $\gamma$ and  $\bm{\kappa}=(0,0,\kappa)$ without losing generality, then define fluctuation amplitudes
 \begin{align}
 \label{rhodef2}
 \rho_n(\omega,\bm{\kappa})\equiv (\tilde{\psi}_n,\tilde{\chi}(\omega,\bm{\kappa},p,\bm{s})\,).
 \end{align}
  Note that the weight function in Eq.(\ref{inner}) is replaced by the equilibrium distribution $f_0=(\frac{z}{2\pi })^{\frac{3}{2}}\exp(-\frac{1}{2}zv^2)$.
 Introduce
$\bm{c}\equiv \sqrt{\frac{z}{2}}\bm{v},\, \bar{c}\equiv\frac{\sqrt{\frac{z}{2}}\omega+i\gamma}{\kappa}, \,\hat{\gamma}\equiv \frac{i\gamma}{\kappa}$,
 and  if $\bar{c}$ is not real, we get
 \begin{align}
 \label{solution1}
 \tilde{\chi}=\frac{\hat{\gamma}\sum_{n=1}^{11}\rho_n\psi_n}{\bar{c}-c_z},
 \end{align}
then these amplitudes can be cast into
 \begin{align}
 \rho_6 &=\hat{\gamma}Z(\bar{c})\rho_6, \nn\\
   \rho_7&=\frac{\hat{\gamma}z^2}{2}\big(\frac{Z(\bar{c})}{z^2}(1+2\bar{c}^2)-\frac{2\bar{c}}{z^2}\,\big)\rho_7+\frac{\hat{\gamma}m}{4\sqrt{V_{9,9}}}\nn\\
   &\times\big(\,(2\bar{c}-\frac{2\bar{c}}{z}-\frac{2\bar{c}^3}{z})Z(\bar{c})+\frac{3}{z}+\frac{2\bar{c}^2}{z}-2\big)\rho_9,\nn\\
   \rho_9&=\frac{\hat{\gamma}m}{4\sqrt{V_{9,9}}}\big(\,(2\bar{c}-\frac{2\bar{c}}{z}-\frac{2\bar{c}^3}{z})Z(\bar{c})+\frac{3}{z}+\frac{2\bar{c}^2}{z}\nn\\
   &-2\big)\rho_7
   +\frac{\hat{\gamma}m^2}{4V_{9,9}}\big(\,(\frac{\bar{c}^4}{z^2}+\frac{3 \bar{c}^2}{2z^2}-\frac{2\bar{c}^2}{z}+\frac{1}{z^2}-\frac{1}{z}\nn\\
   &+1)Z(\bar{c})-\frac{\bar{c}^3}{z^2}-\frac{3\bar{c}}{2z^2}-\frac{\bar{c}}{2z^2}+\frac{2\bar{c}}{z}\big)\rho_{9},\nn\\
   	\rho_{11}& =\frac{\hat{\gamma}m^2}{4V_{9,9}}\big(\,(\frac{\bar{c}^2}{z^2}+\frac{2}{z^2}-\frac{2}{z}+1)Z(\bar{c})-\frac{\bar{c}}{z^2}\,\big)\rho_{11},
 \end{align}
 with $V_{9,9}\equiv\frac{m^2}{4}(1-\frac{2}{z}+\frac{5}{2z^2})$,
  $Z(\bar{c})\equiv\frac{1}{\sqrt{\pi}}\int_{-\infty}^{\infty} dt\exp(-\frac{z}{2}t^2)\frac{1}{\sqrt{\frac{2}{z}}\bar{c}-t}$ and $(\rho_8,\rho_{10})$ is just a copy of $(\rho_7,\rho_9)$. 
  %and $(\rho_8,\rho_{10})$ is a copy of $(\rho_7,\rho_9)$, and not shown either.
 
 The determination of dispersion relations is equivalent to finding roots of 
 \begin{align}
 \Phi_1(\bar{c})&=\hat{\gamma}Z(\bar{c})-1,\nn\\
  \Phi_2(\bar{c},z)&=\frac{\hat{\gamma}m^2}{4V_{9,9}}\big((\frac{\bar{c}^2}{z^2}+\frac{2}{z^2}-\frac{2}{z}+1)Z(\bar{c})-\frac{\bar{c}}{z^2}\big)-1,\nn\\
  \Phi_3(\bar{c},z)&=\big[\frac{\hat{\gamma}z^2}{2}\big(\frac{Z(\bar{c})}{z^2}(1+2\bar{c}^2)-\frac{2\bar{c}}{z^2}\,\big)-1\big]\big[\frac{\hat{\gamma}m^2}{4V_{9,9}}\nn\\
 & \times\big(\,(\frac{\bar{c}^4}{z^2}+\frac{3 \bar{c}^2}{2z^2}-\frac{2\bar{c}^2}{z}+\frac{1}{z^2}-\frac{1}{z}+1)Z(\bar{c})\nn\\
 &-\frac{\bar{c}^3}{z^2}-\frac{3\bar{c}}{2z^2}-\frac{\bar{c}}{2z^2}+\frac{2\bar{c}}{z}\big)-1\big]-\frac{\hat{\gamma}^2m^2}{16V_{9,9}}\nn\\
 &\times\big(\,(2\bar{c}-\frac{2\bar{c}}{z}-\frac{2\bar{c}^3}{z})Z(\bar{c})+\frac{3}{z}+\frac{2\bar{c}^2}{z}-2\big)^2.
 \end{align}
 For illustration we only display how to extract the encoding information from $(\rho_7,\rho_9)$, which can be safely extended to other fluctuation amplitudes. 
 it is convenient to invoke  the residue theorem  that the number of zeros of $\Phi_3$ in a region of the complex $\bar{c}$-plane in which $\Phi_3$ is an analytic function is equal to the times of the representative point  $\Phi_3$ in the $\Phi_3$-plane encircles the origin. 
%A typical illustration of the trajectory of $\Phi_3$ is depicted in Fig.\ref{fig3}, from which one can know whether there are zeros of  $\Phi$ (right) or not (left) by constantly varying the value of $\hat{\gamma}$. 
%\begin{figure}[!htb]
%	\includegraphics[width=0.7\textwidth]{traject} \ ~ \ ~  
%	\caption{The typical trajectories of $\Phi_3(z)$ with $\kappa>\kappa_o$ (left) and $\kappa<\kappa_o$(right)}
%	\label{fig3}
%\end{figure}
 With the asymptotic behavior of $Z(\bar{c})$ for large $|\bar{c}|$ and along the real axis  detailedly given in Appendix.\ref{zc}, it is easy to verify that there is a critical value for $\kappa$ below which zeros exist, i.e. the dispersion relations of discrete normal modes hold, which is also known as an onset of hydrodynamic description \cite{Romatschke:2015gic}. When $\kappa$ exceeds the critical value $\kappa_o$, there are no discrete modes anymore. 
 The critical value $\kappa_o$ for various spin modes are numerically solved and displayed in TABLE.~\ref{table1}. 
\begin{table}[!hbt]
	\begin{tabular}{c|c|c|c}
		\hline\hline
		&\quad L(6th) \quad
		&\quad L(11th) \quad
		& \quad T  \quad
		\\
		\hline
		$\kappa_o/n\sigma\gamma$& 
		1.772 & 
		1.762$\pm$0.040 &
		1.754$\pm$0.079 \\	
		\hline\hline
	\end{tabular}
	\caption{The critical value $\kappa_o$ for longitudinal (L) and transverse (T) spin modes. In the last two columns, $a\pm b$ represents that $a$ is formulated with $z=10$ and $b$ is the maximum discrepancy from $a$ when $z$ ranges from $5$ to $20$.}
	\label{table1}
\end{table}
 The results are exact in the nonrelativistic limit. 
 
 Several comments are followed in order.
 \begin{itemize}
 	\item
 	The present calculation is  limited to nonrelativistic situation, but the qualitative behavior of normal mode should be independent of whether we take this limit.  As the concept of long or short wavelength concerns only the dynamics, it is thus irrelevant to kinematics. We expect  the qualitative behavior of criticality is identical for both relativistic and nonrelativistic cases.
 	%the short or long wavelength behaviors
 	\item
 	The existence of the critical behavior of discrete normal modes reflects the transition from collision-dominated region to Knudsen region, though the realistic distinction may not be  that clear. In simplified mutilated model, the transition region collapses into a critical point and the short or long wavelength dynamics can be uniformly described without extra changes.
 	\item
 	Discrete normal modes including spin modes are exactly the ways of ordered particle collective motion organized by collisions. In long wavelength limit, they  return to zero modes. In free-flow dominated region, the initial fluctuation is carried away by disordered particles. Thus no discrete modes and dispersion relations are found, which happens while $\kappa$ exceeds $\kappa_o$. There is an exception that $\bar{c}$ is real . The behavior of normal modes when $\bar{c}$ is real is alike to  that in Knudsen region.  Two cases are deeply connected with each other by verifying that the fluctuation $\tilde{\chi}$ in both situations takes the form of  single particle continuum spectrum see Appendix.\ref{real1} for details. It makes sense that in the case of long wavelength there are hydrodynamic modes and quasiparticle modes and only  quasiparticle modes are left in Knudsen region.
 	\item
 	As a supplement, the critical $\kappa_o$'s for two sound modes, two shear modes, and one heat mode are $1.853,1.772$ and $1.918$ respectively \cite{deboer}.
 	Thus there exists an another hierarchy $\kappa_{o,\text{spin}}\lesssim \kappa_{o,\text{non}}$, which  manifests that discrete spinless modes are more resistant to the ``destruction'' of non-uniformity. Nevertheless,  this discrepancy is  negligible.
 \end{itemize}

 \section{Summary and outlook}
 \label{su}
 In this paper, we constructed a mutilated model incorporating the collisional invariants in spin Boltzmann equation, where six new zero modes associated with total angular momentum arise from nontrivial dynamics of nonlocal collisions. The dispersion relations of all discrete normal modes, namely, the propagating speeds and attenuation rates are all computed up to second order perturbation. At the first order perturbation in spatial non-uniformity, the non-locality contributes nothing to sound speed but slows down the propagation of spin channel waves. At the second order perturbation, the damping rates of spin modes are close to those of spinless modes for various relevant energy dependence of $\gamma$. The results reveal that spin and momentum relax at comparable damping rates parameterized for modeling realistic  physical scenarios and a unified transportation of both spin and momentum is necessary.  In the nonrelativistic limit, we investigate the existence conditions for discrete normal modes  and find there exists a critical point for every distinct discrete  mode over which only quasiparticle modes contribute.
 
  There are possible extensions to the current work. The first extension lies in a more accurate extraction of parameter $\gamma$ or the reciprocal of relaxation time in RTA. Generally speaking, $\gamma$ is not a free parameter and can be in principle solved from spin Boltzmann equation consistently though it might be complicated to evaluate nonzero eigenvalues and  corresponding eigenfunctions of the full linearized collision operator.  Secondly, our analysis bases itself on the spin Boltzmann equation specialized to the collisions between massive spin-$1/2$ fermions. Therefore, our estimation of  the relaxation time scales for both spin and momentum transport overlooks other relevant processes such as the collisions between massive quark and gluons or massless $u$ and $d$ quarks. It is likely that these scatterings shall play a big role because the rotating media should carry a large amount of orbital angular momentum yielded by non-central nuclear collisions and thus the components,  massless $u$ and $d$ quarks and gluons inevitably polarize or possess considerable orbital angular momentum.  Consequently, gluons and light $u,$ $d$ quarks  are important sources of the polarization of strange quarks as  a result of their mutual interactions.

%\section*{Acknowledgments}
 This work was supported by the NSFC Grant No.11890710, No.11890712 and No.12035006.
%{\bf Acknowledgement:} ...
%--------------------------------------------------------------------------------------------------------------------------------

	\begin{appendix}
	\section{The asymptotic behavior of $Z(\bar{c})$} \label{zc}
	
	In Sec.~\ref{non}, we have introduced an auxiliary function $Z(\bar{c})$ and its asymptotic behavior is crucial to the discussion therein. First we note
	\begin{align}
	\label{zc1}
	Z(\bar{c})\equiv&\frac{1}{\sqrt{\pi}}\int_{-\infty}^{\infty} dt\exp(-\frac{z}{2}t^2)\frac{1}{\sqrt{\frac{2}{z}}\bar{c}-t}\nn\\
	=&\frac{1}{\sqrt{\pi}}\int_{-\infty}^{\infty} d(\sqrt{\frac{z}{2}}t)\exp(-\frac{z}{2}t^2)\frac{1}{\bar{c}-\sqrt{\frac{z}{2}}t}\nn\\
	=&\frac{1}{\sqrt{\pi}}\int_{-\infty}^{\infty} dt\exp(-t^2)\frac{1}{\bar{c}-t}.
	\end{align} 
	For large $|\bar{c}|$, the asymptotic behavior of $Z(\bar{c})$ is
	\begin{align}
	Z(\bar{c})\sim \sum_{n=0}^{\infty}\frac{\Gamma(n+\frac{1}{2})}{\Gamma(\frac{1}{2})}(\bar{c})^{-2n-1},
	\end{align}
	which can be derived by expanding Eq.(\ref{zc1}) in powers of $\bar{c}$ and the behavior of $Z$ along the real axis is given by
	\begin{align}
	\lim\limits_{\epsilon\rightarrow 0}Z(x\pm i\epsilon)=Y(x)\mp i\sqrt{\pi}\exp(-x^2)
	\end{align}
	with $Y(x)$ shown as a Dawson integral
	\begin{align}
	Y(x)=\frac{1}{\sqrt{\pi}}\text{P}\int_{-\infty}^{\infty} dt\,e^{-t^2}\frac{1}{x-t}=2e^{-x^2}\int_{0}^{x}dt\,e^{t^2},
	\end{align}
	where $\rm{P}$ denotes the  Cauchy principal value.

\section{continuum spectrum} \label{real1}

When $\bar{c}$ is real, Eq.~\eqref{solution1}  can not naively be obtained by division.
The formal solution should take the form by noticing $x\delta(x)=0$,
\begin{align}
\label{chi}
&\tilde{\chi}(\omega,\bm{k})=\frac{1}{m}A(\omega,\bm{k})\delta(\bar{c}-c_z)+\hat{\gamma}\rm{P}\frac{\sum_{n=6}^{11}\rho_n\psi_n}{\bar{c}-c_z},
\end{align}
where $\rm{P}$ denotes the  Cauchy principal value.

Defining $\bar{A}_\beta(\omega,\bm{k})\equiv  \frac{1}{2\sqrt{3}\pi}\int d^3\bm{s}\delta(\bm{s}^2-3) A(\omega,\bm{k},\bm{s})\bm{s}_\beta$ and various fluctuation amplitudes can be expressed in terms of two parts, i.e., $\rho_i=\rho_i^{(0)}+\rho_i^{(1)}$,
\begin{align}
\rho^{(1)}_6&=\frac{1}{m\sqrt{V_{6,6}}}(\frac{z}{2\pi })^{\frac{3}{2}}\int d\Gamma \exp(-\frac{z}{2}v^2 )J^{0z}A(\omega,\bm{k},\bm{s})\delta(\bar{c}-c_z)=0,\nn\\
\rho_7^{(1)}&=\frac{1}{m\sqrt{V_{6,6}}}(\frac{z}{2\pi })^{\frac{3}{2}}\int d\Gamma \exp(-\frac{z}{2}v^2 )J^{0x}A(\omega,\bm{k},\bm{s})\delta(\bar{c}-c_z)\nn\\
&=\frac{1}{8\sqrt{\pi}}\bar{A}_y\bar{c}\exp(-\bar{c}^2 ),\nn\\
\rho_9^{(1)}&=\frac{1}{m\sqrt{V^{(0)}_{9,9}}}(\frac{z}{2\pi })^{\frac{3}{2}}\int d\Gamma \exp(-\frac{z}{2}v^2 )
J^{zx}A(\omega,\bm{k},\bm{s})\delta(\bar{c}-c_z)\nn\\
&=\frac{m}{2\pi^{\frac{1}{2}}\sqrt{V^{(0)}_{9,9}}}(1-\frac{\bar{c}^2}{z}-\frac{1}{z})\exp(-\bar{c}^2)\bar{A}_y\nn\\
\rho^{(1)}_{11}&=\frac{1}{m\sqrt{V^{(0)}_{9,9}}}(\frac{z}{2\pi })^{\frac{3}{2}}\int d\Gamma \exp(-\frac{z}{2}v^2 )
J^{xy}A(\omega,\bm{k},\bm{s})\delta(\bar{c}-c_z)\nn\\
&=\frac{m}{2\pi^{\frac{1}{2}}\sqrt{V^{(0)}_{9,9}}}(1-\frac{1}{z})\exp(-\bar{c}^2)\bar{A}_z\nn\\
\end{align} 
and
\begin{align}
\rho^{(0)}_6 &=\hat{\gamma}Y(\bar{c})\rho_6, \nn\\
\rho^{(0)}_7&=\frac{\hat{\gamma}z^2}{2}\big(\frac{Y(\bar{c})}{z^2}(1+2\bar{c}^2)-\frac{2\bar{c}}{z^2}\,\big)\rho_7+\frac{\hat{\gamma}m}{4\sqrt{V^{(0)}_{9,9}}}\big(\,(2\bar{c}-\frac{2\bar{c}}{z}-\frac{2\bar{c}^3}{z})Y(\bar{c})+\frac{3}{z}+\frac{2\bar{c}^2}{z}-2\big)\rho_9,\nn\\
\rho^{(0)}_9&=\frac{\hat{\gamma}m}{4\sqrt{V^{(0)}_{9,9}}}\big(\,(2\bar{c}-\frac{2\bar{c}}{z}-\frac{2\bar{c}^3}{z})Y(\bar{c})+\frac{3}{z}+\frac{2\bar{c}^2}{z}-2\big)\rho_7
+\frac{\hat{\gamma}m^2}{4V^{(0)}_{9,9}}\big(\,(\frac{\bar{c}^4}{z^2}+\frac{3 \bar{c}^2}{2z^2}-\frac{2\bar{c}^2}{z}+\frac{1}{z^2}-\frac{1}{z}\nn\\
&+1)Y(\bar{c})-\frac{\bar{c}^3}{z^2}-\frac{3\bar{c}}{2z^2}-\frac{\bar{c}}{2z^2}+\frac{2\bar{c}}{z}\big)\rho_{9},\nn\\
\rho^{(0)}_{11}& =\frac{\hat{\gamma}m^2}{4V^{(0)}_{9,9}}\big(\,(\frac{\bar{c}^2}{z^2}+\frac{2}{z^2}-\frac{2}{z}+1)Y(\bar{c})-\frac{\bar{c}}{z^2}\,\big)\rho_{11}.
\end{align}
The equation for $\rho_6$ has no roots, which implies that $\rho_6$ is ill-defined in the case of real $\bar{c}$. To fix this problem,  we recombine two longitudinal amplitudes such that $\rho_6^\prime\equiv\frac{1}{\sqrt{2}}(\rho_6+\rho_{11})$, $\rho_{11}^\prime\equiv\frac{1}{\sqrt{2}}(\rho_6-\rho_{11})$. 
%and $\psi^\prime_6\equiv\frac{1}{\sqrt{2}}(\psi_6+\psi_{11}), \psi^\prime_{11}\equiv\frac{1}{\sqrt{2}}(\psi_6-\psi_{11})$.  
Therefore, we get a set of inhomogeneous equations for determining distinct fluctuation amplitudes, while we obtain  a set of homogeneous equations in the case of non-real $\bar{c}$. There won't be a condition requiring vanishing determinant any more. All relevant fluctuation amplitudes are then shown as 
\begin{align}
\label{rho7}
&\rho_7=\rho_8=\frac{-a_4\rho^{(1)}_7+a_2\rho_9^{(1)}}{a_2a_3-a_1a_4},\quad  \rho_9=\rho_{10}=\frac{-a_3\rho^{(1)}_7+a_1\rho_9^{(1)}}{a_2a_3-a_1a_4},\\
\label{rho6}
&\rho^\prime_6=-\rho^\prime_{11}=-\frac{\rho_{11}^{(1)}}{\sqrt{2}}\frac{1}{b-1},
\end{align}
with 
\begin{align}
a_1\equiv&\frac{\hat{\gamma}z^2}{2}\big(\frac{Y(\bar{c})}{z^2}(1+2\bar{c}^2)-\frac{2\bar{c}}{z^2}\,\big)-1,\\
a_2=a_3\equiv &\frac{\hat{\gamma}m}{4\sqrt{V^{(0)}_{9,9}}}\big(\,(2\bar{c}-\frac{2\bar{c}}{z}-\frac{2\bar{c}^3}{z})Y(\bar{c})+\frac{3}{z}+\frac{2\bar{c}^2}{z}-2\big),\\
a_4\equiv&\frac{\hat{\gamma}m^2}{4V^{(0)}_{9,9}}\big(\,(\frac{\bar{c}^4}{z^2}+\frac{3 \bar{c}^2}{2z^2}-\frac{2\bar{c}^2}{z}+\frac{1}{z^2}-\frac{1}{z}+1)Y(\bar{c})-\frac{\bar{c}^3}{z^2}-\frac{3\bar{c}}{2z^2}-\frac{\bar{c}}{2z^2}+\frac{2\bar{c}}{z}\big)-1,\\
b\equiv&\frac{\hat{\gamma}m^2}{4V^{(0)}_{9,9}}\big(\,(\frac{\bar{c}^2}{z^2}+\frac{2}{z^2}-\frac{2}{z}+1)Y(\bar{c})-\frac{\bar{c}}{z^2}\,\big).
\end{align}

From Eqs.~\eqref{rho7} and \eqref{rho6}, we can conclude that six fluctuation amplitudes degenerate into three. Unlike the case of non-real $\bar{c}$ where only discrete spectrum  is allowed (only limited solutions obeying dispersion relations are allowed), almost every $\omega$ and $\bm{k}$ is allowed without further constraint relations. Moreover, all  continuous normal modes are now possessing finite damping rate $\gamma$.
The continuum spectrum represented by Eq.~\eqref{chi} bears resemblance with that in Knudsen region. To show this, one memorizes that free flow effect predominates the effect of the collisions in that region. As a result, the collision term can be safely removed. The eigen spectrum of Eq.(\ref{cf61}) then turns into continuum one-particle spectrum and all eigenfunctions are just shown as delta functions.  It seems that discrete normal modes ($\bar{c}$ is not real) are like some islands surrounded by a sea of continuum modes ($\bar{c}$ is real). When the inhomogeneity ($k$) keeps growing to cross over one after another critical  $k_o$, then the islands are engulfed one by one and eventually there is only a sea left.
	
\end{appendix}

\bibliographystyle{apsrev}
\bibliography{spinmode}{}

\begin{thebibliography}{57}
\expandafter\ifx\csname natexlab\endcsname\relax\def\natexlab#1{#1}\fi
\expandafter\ifx\csname bibnamefont\endcsname\relax
  \def\bibnamefont#1{#1}\fi
\expandafter\ifx\csname bibfnamefont\endcsname\relax
  \def\bibfnamefont#1{#1}\fi
\expandafter\ifx\csname citenamefont\endcsname\relax
  \def\citenamefont#1{#1}\fi
\expandafter\ifx\csname url\endcsname\relax
  \def\url#1{\texttt{#1}}\fi
\expandafter\ifx\csname urlprefix\endcsname\relax\def\urlprefix{URL }\fi
\providecommand{\bibinfo}[2]{#2}
\providecommand{\eprint}[2][]{\url{#2}}

\bibitem[{\citenamefont{Adamczyk et~al.}(2017)}]{STAR:2017ckg}
\bibinfo{author}{\bibfnamefont{L.}~\bibnamefont{Adamczyk}} \bibnamefont{et~al.}
  (\bibinfo{collaboration}{STAR}), \bibinfo{journal}{Nature}
  \textbf{\bibinfo{volume}{548}}, \bibinfo{pages}{62} (\bibinfo{year}{2017}),
  \eprint{1701.06657}.

\bibitem[{\citenamefont{Alpatov}(2020)}]{Alpatov:2020iev}
\bibinfo{author}{\bibfnamefont{E.}~\bibnamefont{Alpatov}}
  (\bibinfo{collaboration}{), STAR (for the}), \bibinfo{journal}{J. Phys. Conf.
  Ser.} \textbf{\bibinfo{volume}{1690}}, \bibinfo{pages}{012120}
  (\bibinfo{year}{2020}).

\bibitem[{\citenamefont{Adam et~al.}(2019)}]{Adam:2019srw}
\bibinfo{author}{\bibfnamefont{J.}~\bibnamefont{Adam}} \bibnamefont{et~al.}
  (\bibinfo{collaboration}{STAR}), \bibinfo{journal}{Phys. Rev. Lett.}
  \textbf{\bibinfo{volume}{123}}, \bibinfo{pages}{132301}
  (\bibinfo{year}{2019}), \eprint{1905.11917}.

\bibitem[{\citenamefont{Adam et~al.}(2018)}]{Adam:2018ivw}
\bibinfo{author}{\bibfnamefont{J.}~\bibnamefont{Adam}} \bibnamefont{et~al.}
  (\bibinfo{collaboration}{STAR}), \bibinfo{journal}{Phys. Rev. C}
  \textbf{\bibinfo{volume}{98}}, \bibinfo{pages}{014910}
  (\bibinfo{year}{2018}), \eprint{1805.04400}.

\bibitem[{\citenamefont{Liang and Wang}(2005)}]{Liang:2004ph}
\bibinfo{author}{\bibfnamefont{Z.-T.} \bibnamefont{Liang}} \bibnamefont{and}
  \bibinfo{author}{\bibfnamefont{X.-N.} \bibnamefont{Wang}},
  \bibinfo{journal}{Phys. Rev. Lett.} \textbf{\bibinfo{volume}{94}},
  \bibinfo{pages}{102301} (\bibinfo{year}{2005}), \bibinfo{note}{[Erratum:
  Phys.Rev.Lett. 96, 039901 (2006)]}, \eprint{nucl-th/0410079}.

\bibitem[{\citenamefont{Wei et~al.}(2019)\citenamefont{Wei, Deng, and
  Huang}}]{Wei:2018zfb}
\bibinfo{author}{\bibfnamefont{D.-X.} \bibnamefont{Wei}},
  \bibinfo{author}{\bibfnamefont{W.-T.} \bibnamefont{Deng}}, \bibnamefont{and}
  \bibinfo{author}{\bibfnamefont{X.-G.} \bibnamefont{Huang}},
  \bibinfo{journal}{Phys. Rev. C} \textbf{\bibinfo{volume}{99}},
  \bibinfo{pages}{014905} (\bibinfo{year}{2019}), \eprint{1810.00151}.

\bibitem[{\citenamefont{Karpenko and Becattini}(2017)}]{Karpenko:2016jyx}
\bibinfo{author}{\bibfnamefont{I.}~\bibnamefont{Karpenko}} \bibnamefont{and}
  \bibinfo{author}{\bibfnamefont{F.}~\bibnamefont{Becattini}},
  \bibinfo{journal}{Eur. Phys. J. C} \textbf{\bibinfo{volume}{77}},
  \bibinfo{pages}{213} (\bibinfo{year}{2017}), \eprint{1610.04717}.

\bibitem[{\citenamefont{Csernai et~al.}(2019)\citenamefont{Csernai, Kapusta,
  and Welle}}]{Csernai:2018yok}
\bibinfo{author}{\bibfnamefont{L.}~\bibnamefont{Csernai}},
  \bibinfo{author}{\bibfnamefont{J.}~\bibnamefont{Kapusta}}, \bibnamefont{and}
  \bibinfo{author}{\bibfnamefont{T.}~\bibnamefont{Welle}},
  \bibinfo{journal}{Phys. Rev. C} \textbf{\bibinfo{volume}{99}},
  \bibinfo{pages}{021901} (\bibinfo{year}{2019}), \eprint{1807.11521}.

\bibitem[{\citenamefont{Bzdak}(2017)}]{Bzdak:2017shg}
\bibinfo{author}{\bibfnamefont{A.}~\bibnamefont{Bzdak}},
  \bibinfo{journal}{Phys. Rev. D} \textbf{\bibinfo{volume}{96}},
  \bibinfo{pages}{056011} (\bibinfo{year}{2017}), \eprint{1703.03003}.

\bibitem[{\citenamefont{Shi et~al.}(2019)\citenamefont{Shi, Li, and
  Liao}}]{Shi:2017wpk}
\bibinfo{author}{\bibfnamefont{S.}~\bibnamefont{Shi}},
  \bibinfo{author}{\bibfnamefont{K.}~\bibnamefont{Li}}, \bibnamefont{and}
  \bibinfo{author}{\bibfnamefont{J.}~\bibnamefont{Liao}},
  \bibinfo{journal}{Phys. Lett. B} \textbf{\bibinfo{volume}{788}},
  \bibinfo{pages}{409} (\bibinfo{year}{2019}), \eprint{1712.00878}.

\bibitem[{\citenamefont{Sun and Ko}(2017)}]{Sun:2017xhx}
\bibinfo{author}{\bibfnamefont{Y.}~\bibnamefont{Sun}} \bibnamefont{and}
  \bibinfo{author}{\bibfnamefont{C.~M.} \bibnamefont{Ko}},
  \bibinfo{journal}{Phys. Rev. C} \textbf{\bibinfo{volume}{96}},
  \bibinfo{pages}{024906} (\bibinfo{year}{2017}), \eprint{1706.09467}.

\bibitem[{\citenamefont{Ivanov et~al.}(2020)\citenamefont{Ivanov, Toneev, and
  Soldatov}}]{Ivanov:2019wzg}
\bibinfo{author}{\bibfnamefont{Y.~B.} \bibnamefont{Ivanov}},
  \bibinfo{author}{\bibfnamefont{V.~D.} \bibnamefont{Toneev}},
  \bibnamefont{and} \bibinfo{author}{\bibfnamefont{A.~A.}
  \bibnamefont{Soldatov}}, \bibinfo{journal}{Phys. Atom. Nucl.}
  \textbf{\bibinfo{volume}{83}}, \bibinfo{pages}{179} (\bibinfo{year}{2020}),
  \eprint{1910.01332}.

\bibitem[{\citenamefont{Xie et~al.}(2017)\citenamefont{Xie, Wang, and
  Csernai}}]{Xie:2017upb}
\bibinfo{author}{\bibfnamefont{Y.}~\bibnamefont{Xie}},
  \bibinfo{author}{\bibfnamefont{D.}~\bibnamefont{Wang}}, \bibnamefont{and}
  \bibinfo{author}{\bibfnamefont{L.~P.} \bibnamefont{Csernai}},
  \bibinfo{journal}{Phys. Rev. C} \textbf{\bibinfo{volume}{95}},
  \bibinfo{pages}{031901} (\bibinfo{year}{2017}), \eprint{1703.03770}.

\bibitem[{\citenamefont{Becattini and Karpenko}(2018)}]{Becattini:2017gcx}
\bibinfo{author}{\bibfnamefont{F.}~\bibnamefont{Becattini}} \bibnamefont{and}
  \bibinfo{author}{\bibfnamefont{I.}~\bibnamefont{Karpenko}},
  \bibinfo{journal}{Phys. Rev. Lett.} \textbf{\bibinfo{volume}{120}},
  \bibinfo{pages}{012302} (\bibinfo{year}{2018}), \eprint{1707.07984}.

\bibitem[{\citenamefont{Xia et~al.}(2018)\citenamefont{Xia, Li, Tang, and
  Wang}}]{Xia:2018tes}
\bibinfo{author}{\bibfnamefont{X.-L.} \bibnamefont{Xia}},
  \bibinfo{author}{\bibfnamefont{H.}~\bibnamefont{Li}},
  \bibinfo{author}{\bibfnamefont{Z.-B.} \bibnamefont{Tang}}, \bibnamefont{and}
  \bibinfo{author}{\bibfnamefont{Q.}~\bibnamefont{Wang}},
  \bibinfo{journal}{Phys. Rev. C} \textbf{\bibinfo{volume}{98}},
  \bibinfo{pages}{024905} (\bibinfo{year}{2018}), \eprint{1803.00867}.

\bibitem[{\citenamefont{Becattini}(2022)}]{Becattini:2022zvf}
\bibinfo{author}{\bibfnamefont{F.}~\bibnamefont{Becattini}}
  (\bibinfo{year}{2022}), \eprint{2204.01144}.

\bibitem[{\citenamefont{Florkowski et~al.}(2018)\citenamefont{Florkowski,
  Friman, Jaiswal, and Speranza}}]{Florkowski:2017ruc}
\bibinfo{author}{\bibfnamefont{W.}~\bibnamefont{Florkowski}},
  \bibinfo{author}{\bibfnamefont{B.}~\bibnamefont{Friman}},
  \bibinfo{author}{\bibfnamefont{A.}~\bibnamefont{Jaiswal}}, \bibnamefont{and}
  \bibinfo{author}{\bibfnamefont{E.}~\bibnamefont{Speranza}},
  \bibinfo{journal}{Phys. Rev. C} \textbf{\bibinfo{volume}{97}},
  \bibinfo{pages}{041901} (\bibinfo{year}{2018}), \eprint{1705.00587}.

\bibitem[{\citenamefont{Peng et~al.}(2021)\citenamefont{Peng, Zhang, Sheng, and
  Wang}}]{Peng:2021ago}
\bibinfo{author}{\bibfnamefont{H.-H.} \bibnamefont{Peng}},
  \bibinfo{author}{\bibfnamefont{J.-J.} \bibnamefont{Zhang}},
  \bibinfo{author}{\bibfnamefont{X.-L.} \bibnamefont{Sheng}}, \bibnamefont{and}
  \bibinfo{author}{\bibfnamefont{Q.}~\bibnamefont{Wang}}
  (\bibinfo{year}{2021}), \eprint{2107.00448}.

\bibitem[{\citenamefont{Becattini and Tinti}(2010)}]{Becattini:2009wh}
\bibinfo{author}{\bibfnamefont{F.}~\bibnamefont{Becattini}} \bibnamefont{and}
  \bibinfo{author}{\bibfnamefont{L.}~\bibnamefont{Tinti}},
  \bibinfo{journal}{Annals Phys.} \textbf{\bibinfo{volume}{325}},
  \bibinfo{pages}{1566} (\bibinfo{year}{2010}), \eprint{0911.0864}.

\bibitem[{\citenamefont{Florkowski et~al.}(2019)\citenamefont{Florkowski,
  Kumar, and Ryblewski}}]{Florkowski:2018fap}
\bibinfo{author}{\bibfnamefont{W.}~\bibnamefont{Florkowski}},
  \bibinfo{author}{\bibfnamefont{A.}~\bibnamefont{Kumar}}, \bibnamefont{and}
  \bibinfo{author}{\bibfnamefont{R.}~\bibnamefont{Ryblewski}},
  \bibinfo{journal}{Prog. Part. Nucl. Phys.} \textbf{\bibinfo{volume}{108}},
  \bibinfo{pages}{103709} (\bibinfo{year}{2019}), \eprint{1811.04409}.

\bibitem[{\citenamefont{Hattori et~al.}(2019)\citenamefont{Hattori, Hongo,
  Huang, Matsuo, and Taya}}]{Hattori:2019lfp}
\bibinfo{author}{\bibfnamefont{K.}~\bibnamefont{Hattori}},
  \bibinfo{author}{\bibfnamefont{M.}~\bibnamefont{Hongo}},
  \bibinfo{author}{\bibfnamefont{X.-G.} \bibnamefont{Huang}},
  \bibinfo{author}{\bibfnamefont{M.}~\bibnamefont{Matsuo}}, \bibnamefont{and}
  \bibinfo{author}{\bibfnamefont{H.}~\bibnamefont{Taya}},
  \bibinfo{journal}{Phys. Lett. B} \textbf{\bibinfo{volume}{795}},
  \bibinfo{pages}{100} (\bibinfo{year}{2019}), \eprint{1901.06615}.

\bibitem[{\citenamefont{Fukushima and Pu}(2021)}]{Fukushima:2020ucl}
\bibinfo{author}{\bibfnamefont{K.}~\bibnamefont{Fukushima}} \bibnamefont{and}
  \bibinfo{author}{\bibfnamefont{S.}~\bibnamefont{Pu}}, \bibinfo{journal}{Phys.
  Lett. B} \textbf{\bibinfo{volume}{817}}, \bibinfo{pages}{136346}
  (\bibinfo{year}{2021}), \eprint{2010.01608}.

\bibitem[{\citenamefont{Hu}(2021)}]{Hu:2021lnx}
\bibinfo{author}{\bibfnamefont{J.}~\bibnamefont{Hu}}, \bibinfo{journal}{Phys.
  Rev. D} \textbf{\bibinfo{volume}{103}}, \bibinfo{pages}{116015}
  (\bibinfo{year}{2021}), \eprint{2101.08440}.

\bibitem[{\citenamefont{Bhadury et~al.}(2021)\citenamefont{Bhadury, Florkowski,
  Jaiswal, Kumar, and Ryblewski}}]{Bhadury:2020cop}
\bibinfo{author}{\bibfnamefont{S.}~\bibnamefont{Bhadury}},
  \bibinfo{author}{\bibfnamefont{W.}~\bibnamefont{Florkowski}},
  \bibinfo{author}{\bibfnamefont{A.}~\bibnamefont{Jaiswal}},
  \bibinfo{author}{\bibfnamefont{A.}~\bibnamefont{Kumar}}, \bibnamefont{and}
  \bibinfo{author}{\bibfnamefont{R.}~\bibnamefont{Ryblewski}},
  \bibinfo{journal}{Phys. Rev. D} \textbf{\bibinfo{volume}{103}},
  \bibinfo{pages}{014030} (\bibinfo{year}{2021}), \eprint{2008.10976}.

\bibitem[{\citenamefont{Hu}(2022{\natexlab{a}})}]{Hu:2021pwh}
\bibinfo{author}{\bibfnamefont{J.}~\bibnamefont{Hu}}, \bibinfo{journal}{Phys.
  Rev. D} \textbf{\bibinfo{volume}{105}}, \bibinfo{pages}{076009}
  (\bibinfo{year}{2022}{\natexlab{a}}), \eprint{2111.03571}.

\bibitem[{\citenamefont{Hu}(2022{\natexlab{b}})}]{Hu:2022lpi}
\bibinfo{author}{\bibfnamefont{J.}~\bibnamefont{Hu}}, \bibinfo{journal}{Phys.
  Rev. D} \textbf{\bibinfo{volume}{106}}, \bibinfo{pages}{036004}
  (\bibinfo{year}{2022}{\natexlab{b}}), \eprint{2202.07373}.

\bibitem[{\citenamefont{Hu}(2022{\natexlab{c}})}]{Hu:2022xjn}
\bibinfo{author}{\bibfnamefont{J.}~\bibnamefont{Hu}}, \bibinfo{journal}{Phys.
  Rev. D} \textbf{\bibinfo{volume}{105}}, \bibinfo{pages}{096021}
  (\bibinfo{year}{2022}{\natexlab{c}}), \eprint{2204.12946}.

\bibitem[{\citenamefont{Hu}(2022{\natexlab{d}})}]{Hu:2022azy}
\bibinfo{author}{\bibfnamefont{J.}~\bibnamefont{Hu}}
  (\bibinfo{year}{2022}{\natexlab{d}}), \eprint{2209.10979}.

\bibitem[{\citenamefont{Weickgenannt et~al.}(2022)\citenamefont{Weickgenannt,
  Wagner, Speranza, and Rischke}}]{Weickgenannt:2022zxs}
\bibinfo{author}{\bibfnamefont{N.}~\bibnamefont{Weickgenannt}},
  \bibinfo{author}{\bibfnamefont{D.}~\bibnamefont{Wagner}},
  \bibinfo{author}{\bibfnamefont{E.}~\bibnamefont{Speranza}}, \bibnamefont{and}
  \bibinfo{author}{\bibfnamefont{D.}~\bibnamefont{Rischke}}
  (\bibinfo{year}{2022}), \eprint{2203.04766}.

\bibitem[{\citenamefont{Weickgenannt
  et~al.}(2021{\natexlab{a}})\citenamefont{Weickgenannt, Speranza, Sheng, Wang,
  and Rischke}}]{Weickgenannt:2021cuo}
\bibinfo{author}{\bibfnamefont{N.}~\bibnamefont{Weickgenannt}},
  \bibinfo{author}{\bibfnamefont{E.}~\bibnamefont{Speranza}},
  \bibinfo{author}{\bibfnamefont{X.-l.} \bibnamefont{Sheng}},
  \bibinfo{author}{\bibfnamefont{Q.}~\bibnamefont{Wang}}, \bibnamefont{and}
  \bibinfo{author}{\bibfnamefont{D.~H.} \bibnamefont{Rischke}},
  \bibinfo{journal}{Phys. Rev. D} \textbf{\bibinfo{volume}{104}},
  \bibinfo{pages}{016022} (\bibinfo{year}{2021}{\natexlab{a}}),
  \eprint{2103.04896}.

\bibitem[{\citenamefont{Weickgenannt
  et~al.}(2021{\natexlab{b}})\citenamefont{Weickgenannt, Speranza, Sheng, Wang,
  and Rischke}}]{Weickgenannt:2020aaf}
\bibinfo{author}{\bibfnamefont{N.}~\bibnamefont{Weickgenannt}},
  \bibinfo{author}{\bibfnamefont{E.}~\bibnamefont{Speranza}},
  \bibinfo{author}{\bibfnamefont{X.-l.} \bibnamefont{Sheng}},
  \bibinfo{author}{\bibfnamefont{Q.}~\bibnamefont{Wang}}, \bibnamefont{and}
  \bibinfo{author}{\bibfnamefont{D.~H.} \bibnamefont{Rischke}},
  \bibinfo{journal}{Phys. Rev. Lett.} \textbf{\bibinfo{volume}{127}},
  \bibinfo{pages}{052301} (\bibinfo{year}{2021}{\natexlab{b}}),
  \eprint{2005.01506}.

\bibitem[{\citenamefont{Yang et~al.}(2020)\citenamefont{Yang, Hattori, and
  Hidaka}}]{Yang:2020hri}
\bibinfo{author}{\bibfnamefont{D.-L.} \bibnamefont{Yang}},
  \bibinfo{author}{\bibfnamefont{K.}~\bibnamefont{Hattori}}, \bibnamefont{and}
  \bibinfo{author}{\bibfnamefont{Y.}~\bibnamefont{Hidaka}},
  \bibinfo{journal}{JHEP} \textbf{\bibinfo{volume}{20}}, \bibinfo{pages}{070}
  (\bibinfo{year}{2020}), \eprint{2002.02612}.

\bibitem[{\citenamefont{Sheng et~al.}(2021)\citenamefont{Sheng, Weickgenannt,
  Speranza, Rischke, and Wang}}]{Sheng:2021kfc}
\bibinfo{author}{\bibfnamefont{X.-L.} \bibnamefont{Sheng}},
  \bibinfo{author}{\bibfnamefont{N.}~\bibnamefont{Weickgenannt}},
  \bibinfo{author}{\bibfnamefont{E.}~\bibnamefont{Speranza}},
  \bibinfo{author}{\bibfnamefont{D.~H.} \bibnamefont{Rischke}},
  \bibnamefont{and} \bibinfo{author}{\bibfnamefont{Q.}~\bibnamefont{Wang}},
  \bibinfo{journal}{Phys. Rev. D} \textbf{\bibinfo{volume}{104}},
  \bibinfo{pages}{016029} (\bibinfo{year}{2021}), \eprint{2103.10636}.

\bibitem[{\citenamefont{Chen and Lin}(2022)}]{Chen:2021azy}
\bibinfo{author}{\bibfnamefont{Z.}~\bibnamefont{Chen}} \bibnamefont{and}
  \bibinfo{author}{\bibfnamefont{S.}~\bibnamefont{Lin}},
  \bibinfo{journal}{Phys. Rev. D} \textbf{\bibinfo{volume}{105}},
  \bibinfo{pages}{014015} (\bibinfo{year}{2022}), \eprint{2109.08440}.

\bibitem[{\citenamefont{Wang and Zhuang}(2021)}]{Wang:2021qnt}
\bibinfo{author}{\bibfnamefont{Z.}~\bibnamefont{Wang}} \bibnamefont{and}
  \bibinfo{author}{\bibfnamefont{P.}~\bibnamefont{Zhuang}}
  (\bibinfo{year}{2021}), \eprint{2105.00915}.

\bibitem[{\citenamefont{Yang}(2021)}]{Yang:2021fea}
\bibinfo{author}{\bibfnamefont{D.-L.} \bibnamefont{Yang}}
  (\bibinfo{year}{2021}), \eprint{2112.14392}.

\bibitem[{\citenamefont{Becattini et~al.}(2019)\citenamefont{Becattini,
  Florkowski, and Speranza}}]{Becattini:2018duy}
\bibinfo{author}{\bibfnamefont{F.}~\bibnamefont{Becattini}},
  \bibinfo{author}{\bibfnamefont{W.}~\bibnamefont{Florkowski}},
  \bibnamefont{and} \bibinfo{author}{\bibfnamefont{E.}~\bibnamefont{Speranza}},
  \bibinfo{journal}{Phys. Lett. B} \textbf{\bibinfo{volume}{789}},
  \bibinfo{pages}{419} (\bibinfo{year}{2019}), \eprint{1807.10994}.

\bibitem[{\citenamefont{Li and Yee}(2019)}]{Li:2019qkf}
\bibinfo{author}{\bibfnamefont{S.}~\bibnamefont{Li}} \bibnamefont{and}
  \bibinfo{author}{\bibfnamefont{H.-U.} \bibnamefont{Yee}},
  \bibinfo{journal}{Phys. Rev. D} \textbf{\bibinfo{volume}{100}},
  \bibinfo{pages}{056022} (\bibinfo{year}{2019}), \eprint{1905.10463}.

\bibitem[{\citenamefont{Kapusta
  et~al.}(2020{\natexlab{a}})\citenamefont{Kapusta, Rrapaj, and
  Rudaz}}]{Kapusta:2019sad}
\bibinfo{author}{\bibfnamefont{J.~I.} \bibnamefont{Kapusta}},
  \bibinfo{author}{\bibfnamefont{E.}~\bibnamefont{Rrapaj}}, \bibnamefont{and}
  \bibinfo{author}{\bibfnamefont{S.}~\bibnamefont{Rudaz}},
  \bibinfo{journal}{Phys. Rev. C} \textbf{\bibinfo{volume}{101}},
  \bibinfo{pages}{024907} (\bibinfo{year}{2020}{\natexlab{a}}),
  \eprint{1907.10750}.

\bibitem[{\citenamefont{Kapusta
  et~al.}(2020{\natexlab{b}})\citenamefont{Kapusta, Rrapaj, and
  Rudaz}}]{Kapusta:2020npk}
\bibinfo{author}{\bibfnamefont{J.~I.} \bibnamefont{Kapusta}},
  \bibinfo{author}{\bibfnamefont{E.}~\bibnamefont{Rrapaj}}, \bibnamefont{and}
  \bibinfo{author}{\bibfnamefont{S.}~\bibnamefont{Rudaz}},
  \bibinfo{journal}{Phys. Rev. C} \textbf{\bibinfo{volume}{102}},
  \bibinfo{pages}{064911} (\bibinfo{year}{2020}{\natexlab{b}}),
  \eprint{2004.14807}.

\bibitem[{\citenamefont{Hongo et~al.}(2022)\citenamefont{Hongo, Huang,
  Kaminski, Stephanov, and Yee}}]{Hongo:2022izs}
\bibinfo{author}{\bibfnamefont{M.}~\bibnamefont{Hongo}},
  \bibinfo{author}{\bibfnamefont{X.-G.} \bibnamefont{Huang}},
  \bibinfo{author}{\bibfnamefont{M.}~\bibnamefont{Kaminski}},
  \bibinfo{author}{\bibfnamefont{M.}~\bibnamefont{Stephanov}},
  \bibnamefont{and} \bibinfo{author}{\bibfnamefont{H.-U.} \bibnamefont{Yee}}
  (\bibinfo{year}{2022}), \eprint{2201.12390}.

\bibitem[{\citenamefont{Kapusta
  et~al.}(2020{\natexlab{c}})\citenamefont{Kapusta, Rrapaj, and
  Rudaz}}]{Kapusta:2019ktm}
\bibinfo{author}{\bibfnamefont{J.~I.} \bibnamefont{Kapusta}},
  \bibinfo{author}{\bibfnamefont{E.}~\bibnamefont{Rrapaj}}, \bibnamefont{and}
  \bibinfo{author}{\bibfnamefont{S.}~\bibnamefont{Rudaz}},
  \bibinfo{journal}{Phys. Rev. C} \textbf{\bibinfo{volume}{101}},
  \bibinfo{pages}{031901} (\bibinfo{year}{2020}{\natexlab{c}}),
  \eprint{1910.12759}.

\bibitem[{\citenamefont{Li and Yee}(2018)}]{Li:2018srq}
\bibinfo{author}{\bibfnamefont{S.}~\bibnamefont{Li}} \bibnamefont{and}
  \bibinfo{author}{\bibfnamefont{H.-U.} \bibnamefont{Yee}},
  \bibinfo{journal}{Phys. Rev. D} \textbf{\bibinfo{volume}{98}},
  \bibinfo{pages}{056018} (\bibinfo{year}{2018}), \eprint{1805.04057}.

\bibitem[{\citenamefont{Ayala et~al.}(2020{\natexlab{a}})\citenamefont{Ayala,
  De~La~Cruz, Hern\'andez-Ort\'\i{}z, Hern\'andez, and
  Salinas}}]{Ayala:2019iin}
\bibinfo{author}{\bibfnamefont{A.}~\bibnamefont{Ayala}},
  \bibinfo{author}{\bibfnamefont{D.}~\bibnamefont{De~La~Cruz}},
  \bibinfo{author}{\bibfnamefont{S.}~\bibnamefont{Hern\'andez-Ort\'\i{}z}},
  \bibinfo{author}{\bibfnamefont{L.~A.} \bibnamefont{Hern\'andez}},
  \bibnamefont{and} \bibinfo{author}{\bibfnamefont{J.}~\bibnamefont{Salinas}},
  \bibinfo{journal}{Phys. Lett. B} \textbf{\bibinfo{volume}{801}},
  \bibinfo{pages}{135169} (\bibinfo{year}{2020}{\natexlab{a}}),
  \eprint{1909.00274}.

\bibitem[{\citenamefont{Ayala et~al.}(2020{\natexlab{b}})\citenamefont{Ayala,
  de~la Cruz, Hern\'andez, and Salinas}}]{Ayala:2020ndx}
\bibinfo{author}{\bibfnamefont{A.}~\bibnamefont{Ayala}},
  \bibinfo{author}{\bibfnamefont{D.}~\bibnamefont{de~la Cruz}},
  \bibinfo{author}{\bibfnamefont{L.~A.} \bibnamefont{Hern\'andez}},
  \bibnamefont{and} \bibinfo{author}{\bibfnamefont{J.}~\bibnamefont{Salinas}},
  \bibinfo{journal}{Phys. Rev. D} \textbf{\bibinfo{volume}{102}},
  \bibinfo{pages}{056019} (\bibinfo{year}{2020}{\natexlab{b}}),
  \eprint{2003.06545}.

\bibitem[{\citenamefont{De~Groot et~al.}(1980)\citenamefont{De~Groot,
  Van~Leeuwen, and Van~Weert}}]{DeGroot:1980dk}
\bibinfo{author}{\bibfnamefont{S.~R.} \bibnamefont{De~Groot}},
  \bibinfo{author}{\bibfnamefont{W.~A.} \bibnamefont{Van~Leeuwen}},
  \bibnamefont{and} \bibinfo{author}{\bibfnamefont{C.~G.}
  \bibnamefont{Van~Weert}}, \emph{\bibinfo{title}{Relativistic Kinetic Theory.
  Principles and Applications}} (\bibinfo{publisher}{North-Holland},
  \bibinfo{year}{1980}).

\bibitem[{\citenamefont{Hongo et~al.}(2021)\citenamefont{Hongo, Huang,
  Kaminski, Stephanov, and Yee}}]{Hongo:2021ona}
\bibinfo{author}{\bibfnamefont{M.}~\bibnamefont{Hongo}},
  \bibinfo{author}{\bibfnamefont{X.-G.} \bibnamefont{Huang}},
  \bibinfo{author}{\bibfnamefont{M.}~\bibnamefont{Kaminski}},
  \bibinfo{author}{\bibfnamefont{M.}~\bibnamefont{Stephanov}},
  \bibnamefont{and} \bibinfo{author}{\bibfnamefont{H.-U.} \bibnamefont{Yee}},
  \bibinfo{journal}{JHEP} \textbf{\bibinfo{volume}{11}}, \bibinfo{pages}{150}
  (\bibinfo{year}{2021}), \eprint{2107.14231}.

\bibitem[{\citenamefont{Rocha et~al.}(2021)\citenamefont{Rocha, Denicol, and
  Noronha}}]{Rocha:2021zcw}
\bibinfo{author}{\bibfnamefont{G.~S.} \bibnamefont{Rocha}},
  \bibinfo{author}{\bibfnamefont{G.~S.} \bibnamefont{Denicol}},
  \bibnamefont{and} \bibinfo{author}{\bibfnamefont{J.}~\bibnamefont{Noronha}},
  \bibinfo{journal}{Phys. Rev. Lett.} \textbf{\bibinfo{volume}{127}},
  \bibinfo{pages}{042301} (\bibinfo{year}{2021}), \eprint{2103.07489}.

\bibitem[{\citenamefont{Ambrus et~al.}(2022)\citenamefont{Ambrus, Ryblewski,
  and Singh}}]{Ambrus:2022yzz}
\bibinfo{author}{\bibfnamefont{V.~E.} \bibnamefont{Ambrus}},
  \bibinfo{author}{\bibfnamefont{R.}~\bibnamefont{Ryblewski}},
  \bibnamefont{and} \bibinfo{author}{\bibfnamefont{R.}~\bibnamefont{Singh}}
  (\bibinfo{year}{2022}), \eprint{2202.03952}.

\bibitem[{\citenamefont{Malfliet}(1984)}]{Malfliet:1984ix}
\bibinfo{author}{\bibfnamefont{R.}~\bibnamefont{Malfliet}},
  \bibinfo{journal}{Nucl. Phys. A} \textbf{\bibinfo{volume}{420}},
  \bibinfo{pages}{621} (\bibinfo{year}{1984}).

\bibitem[{\citenamefont{{Anderson} and {Witting}}(1974)}]{1974Phy....74..466A}
\bibinfo{author}{\bibfnamefont{J.~L.} \bibnamefont{{Anderson}}}
  \bibnamefont{and} \bibinfo{author}{\bibfnamefont{H.~R.}
  \bibnamefont{{Witting}}}, \bibinfo{journal}{Physica}
  \textbf{\bibinfo{volume}{74}}, \bibinfo{pages}{466} (\bibinfo{year}{1974}).

\bibitem[{\citenamefont{Dusling et~al.}(2010)\citenamefont{Dusling, Moore, and
  Teaney}}]{Dusling:2009df}
\bibinfo{author}{\bibfnamefont{K.}~\bibnamefont{Dusling}},
  \bibinfo{author}{\bibfnamefont{G.~D.} \bibnamefont{Moore}}, \bibnamefont{and}
  \bibinfo{author}{\bibfnamefont{D.}~\bibnamefont{Teaney}},
  \bibinfo{journal}{Phys. Rev. C} \textbf{\bibinfo{volume}{81}},
  \bibinfo{pages}{034907} (\bibinfo{year}{2010}), \eprint{0909.0754}.

\bibitem[{\citenamefont{Dusling and Sch\"afer}(2012)}]{Dusling:2011fd}
\bibinfo{author}{\bibfnamefont{K.}~\bibnamefont{Dusling}} \bibnamefont{and}
  \bibinfo{author}{\bibfnamefont{T.}~\bibnamefont{Sch\"afer}},
  \bibinfo{journal}{Phys. Rev. C} \textbf{\bibinfo{volume}{85}},
  \bibinfo{pages}{044909} (\bibinfo{year}{2012}), \eprint{1109.5181}.

\bibitem[{\citenamefont{Kurkela and Wiedemann}(2019)}]{Kurkela:2017xis}
\bibinfo{author}{\bibfnamefont{A.}~\bibnamefont{Kurkela}} \bibnamefont{and}
  \bibinfo{author}{\bibfnamefont{U.~A.} \bibnamefont{Wiedemann}},
  \bibinfo{journal}{Eur. Phys. J. C} \textbf{\bibinfo{volume}{79}},
  \bibinfo{pages}{776} (\bibinfo{year}{2019}), \eprint{1712.04376}.

\bibitem[{\citenamefont{Balescu}(1975)}]{Balescu}
\bibinfo{author}{\bibfnamefont{R.}~\bibnamefont{Balescu}},
  \emph{\bibinfo{title}{Equilibrium and Nonequilibrium Statistical Mechanics}}
  (\bibinfo{publisher}{A Wiley-Interscience}, \bibinfo{year}{1975}).

\bibitem[{\citenamefont{Romatschke}(2016)}]{Romatschke:2015gic}
\bibinfo{author}{\bibfnamefont{P.}~\bibnamefont{Romatschke}},
  \bibinfo{journal}{Eur. Phys. J. C} \textbf{\bibinfo{volume}{76}},
  \bibinfo{pages}{352} (\bibinfo{year}{2016}), \eprint{1512.02641}.

\bibitem[{\citenamefont{Boer and G.E.Uhlenbeck}(1970)}]{deboer}
\bibinfo{author}{\bibfnamefont{J.}~\bibnamefont{Boer}} \bibnamefont{and}
  \bibinfo{author}{\bibnamefont{G.E.Uhlenbeck}}, \emph{\bibinfo{title}{Studies
  in the Statistical Mechanics}} (\bibinfo{publisher}{Orth-Holland Publishing
  Company, Amsterdam}, \bibinfo{year}{1970}).

\end{thebibliography}

\end{document}